\documentclass[12pt,a4paper]{article}

\usepackage{amsmath}
\usepackage{graphicx}
\usepackage{times}
\usepackage{cite}
\begin{document}
\begin{titlepage}
\title{Reflective scattering, color conductivity, and centrality in  hadron reactions}
\author{ S.M. Troshin, N.E. Tyurin\\[1ex]
\small  \it NRC ``Kurchatov Institute''--IHEP\\
\small  \it Protvino, 142281, Russian Federation,\\
\small Sergey.Troshin@ihep.ru
}
\normalsize
\date{}
\maketitle

\begin{abstract}
 The reflective scattering mode is supposed to play a significant role in hadron interactions at the LHC energy region and beyond.  We discuss  connection of this mode to the color conducting medium formation in hadron collisions and its role in centrality determination.  The issues
of centrality   in view of the  measurements at the LHC  are relevant for enlightening the asymptotic   dynamics.
\end{abstract}
\end{titlepage}
\setcounter{page}{2}
\section{Introduction}
Hadrons are  composite, extended objects and their formfactors are  described by nontrivial functions. So,   a seemingly natural expectation  is an increase with energy of the relative weight  of all the inelastic interactions. Otherwise,  a significantly increasing relative contribution to $pp$--interactions  from  elastic scattering events was observed, i.e. the ratio of  elastic to total cross-sections $\sigma_{el}(s)/\sigma_{tot}(s)$  rises with  energy, while the ratio $\sigma_{inel}(s)/\sigma_{tot}(s)$ decreases accordingly. Such trends in behavior of elastic and inelastic interactions have been  confirmed by the recent measurements at $\sqrt{s}=13$ TeV \cite {rat13}. Thus, the experimental results are divergent from the above  expectation which  looked as a natural one for the time being. However, such divergence between experimental results and theoretical expectation is not surprising due to neglect of the  color confinement effects under formulation of such naive suggestion.  Evidently, the role of confinement becomes more significant as the hadron collision energy increases. 

An important role of the elastic component of strong interactions dynamics  was underlined long time ago by Chew and Frautchi \cite{chew} in their study based on the Mandelstam representation and principle of maximal strength of strong interaction. The term ``strongest possible" interaction means maximality of the {\it total} interaction rate $N_{tot}$ which is a product of the collider luminosity by the effective total cross--section,  i.e. $N_{tot}={\cal {L}} \sigma _{tot}$.  It corresponds  to maximal possible values of {\it imaginary}  parts of the partial amplitudes since sum of those results in an effective total cross-section according to optical theorem. Chew and Frautchi  noted that a ``{\it characteristic of strong interactions is a capacity to ``saturate'' the unitarity condition at high energies}''.  It results in  a hint on the dominating contribution of the elastic scattering ({\it decoupled from the multiparticle production })   due to unitarity saturation at $s\to\infty$. Unitarity saturation corresponds  to {\it zeroing of the real part} of partial amplitudes\footnote{It is evident from the Argand plot for a partial amplitude.}.

 It should also be remarked here (enhancing what was said  above),  that the upper bound for the inelastic cross--section obtained recently  \cite{wu} {\it excludes}  $\ln^2 s$-dependence for $\sigma_{inel}(s)$ at $s\to\infty$ when the ratio
 $\sigma_{tot}(s)/\sigma_{tot}^{max}(s)$ follows its limiting behavior, i.e. it tends to unity at $s\to\infty$. Here $\sigma_{tot}^{max}(s)$ ($ \equiv (4\pi/{t_0})\ln^2(s/s_0)$),  represents the Froissart--Martin boundary dependence for the total cross-sections and results from saturation of unitarity  and the elastic scattering amplitude analyticity  in the Lehmann-Martin ellipse.
 
Extrapolation of the observed experimental dependencies to  higher energies can be performed in a twofold  way: one can assume equipartition of the elastic and inelastic contributions at $s\to\infty$ (black disc) or  saturation of the unitarity bound leading to the asymptotic behavior of the ratios $\sigma_{el}(s)/\sigma_{tot}(s)\to 1$ and $\sigma_{inel}(s)/\sigma_{tot}(s)\to 0$. Some intermediate dependencies of these ratios at $s\to \infty$ can also be envisaged (cf.  \cite{fagund}). Both of the above mentioned modes are in agreement with unitarity, but rejection of the reflective scattering by assuming $f(s,b)\leq 1/2$, where $b$ is an impact parameter of the colliding hadrons (note that $l=b\sqrt{s}/2$ and the real part of the amplitude is neglected) means {\it ad hoc} limitation imposed by  assumption that the only possible scattering mode it is ``the shadow mode'' and  restricts the wealth of the possible dynamics provided by the unitarity.  Moreover,  it has been known for a long time that absorption is not a consequence of unitarity \cite{sachr}.

The reflective scattering mode corresponds to unitarity saturation at $s\to\infty$, decoupling of elastic scattering from multiparticle production in this limit and, hence, is  consistent with  ``strip approximation'' of \cite{chew} which suppose dominance of double spectral functions in the areas determined by elastic unitarity condition. The main feature of the  mode is a negative value  of the elastic scattering matrix element $S(s,b)$, it leads to   asymptotic dominance of the elastic scattering  and peripheral character  of  the inelastic scattering overlap function in the impact parameter space.
Decoupling of  the elastic scattering  from the multiparticle production starts to occur  at small values of the impact parameter $b$ first and expands to larger values of $b$ while collision energy increases. Such a behavior corresponds to increasing self--dumping of  inelastic contributions to unitarity equation \cite{baker}. The knowledge of the decoupling dynamics   is essential   for  the hadron interaction's studies, e.g. for the  development of QCD in its nonperturbative sector where the color confinement plays an essential role. 

The aforementioned properties of  the reflective scattering  underlines importance of this mode  in the context of the hadron dynamics study. The interest  is actualized also by the results of the recent measurements at the LHC \cite{difs, difs1}.  The essential features of the reflective scattering mode are discussed in the  first part of this note.

The $b$--dependence of the scattering amplitude as well as of the inelastic overlap function are associated with  a collision geometry. The quantitative description of  these functions is important for   the  reflective scattering mode detection   occurring initially at small impact parameter values. This can be performed  using centrality variable for the events classification. We consider suggestion for centrality definition in the case of small systems in the second part of this note.

\section{Reflective scattering in hadron interaction}
The main features of the reflective scattering mode are listed below. Partial wave matrix element of the elastic scattering  is related to the corresponding amplitude $f_l(s)$ by the relation 
\[
S_l(s)=1+2if_l(s),
\]
where the amplitude $f_l(s)$ obey the unitarity equation:
 \begin{equation}
 \mbox{Im} f_l(s)=|f_l(s)|^2+h_{l,{inel}}(s). \label{ub}
 \end{equation}
It is convenient to use an impact parameter representation, which 
 provides a simple semiclassical picture of hadron scattering, recall that $l=b\sqrt{s}/2$.
For simplicity, we use a common assumption on the smallness of the
 real part of the elastic scattering amplitude  in the impact parameter representation $f(s,b)$  and perform replacement $f\to if$. It should be noted that this assumption correlates with unitarity saturation. Indeed,  unitarity saturation means that $\mbox{Im} f(s,b)\to 1$ at $s\to\infty$ and fixed $b$. It can easily be seen that this limiting behavior implies that  $\mbox{Re} f(s,b)\to 0$ at $s\to\infty$ and fixed $b$ and this leads to inconsistency \cite{pl} of Maximal Odderon \cite{nic} with unitarity saturation.  Neglect of the real part contribution is  justified at least qualitatively since the  recent  analysis \cite{alkin1} is consistent with this conclusion. However, the problem of the  real part  cannot be considered as solved nowadays due to insufficient experimental data. Its value is sensitive to the possible violation of dispersion relations \cite{khuri}. 
 It should be noted that there were considered different amplitude parameterizations in \cite{alkin1}. Those include the ones with {\it negative } imaginary part at large values of $-t$ and all are consistent, nonetheless, with peripheral form of the inelastic overlap function. For discussion of the imaginary part of the scattering amplitude sign, see \cite{drem}.

Unitarity equation provides the evident relation for the dimensionless differential distribution  of the inelastic collisions over $b$ $h_{inel}(s,b)$  in  case of proton--proton scattering
\begin{equation}\label{pinel}
h_{inel}(s,b)= f(s,b)(1-f(s,b)).
\end{equation}
It  constraints variation of the  amplitude $f(s,b)$ by the values from the interval
$0 \leq f \leq 1$. The value of $f=1/2$ corresponds to the complete absorption of the initial state and means that the elastic scattering matrix element  is zero, $S=0$  (note that $S=1-2f$).  If the amplitude $f(s,b)$ at $b=0$ (beyond some threshold value of energy) becomes greater than $1/2$,  then the maximal value of differential distribution of inelastic collisions is $1/4$ at $b> 0$ (cf. Eq. (\ref{pinel})).  Thus, approaching unitarity saturation limit in the region where $f>1/2$ leads to a peripheral nature of the inelastic hadron collisions' differential distribution over impact parameter (inelastic overlap function). Such peripheral character is a straightforward result of a probability conservation, i.e. unitarity.

 Reflective scattering mode appears in the region of $s$ and $b$ where the amplitude $f$ variates in the range $1/2 < f \leq 1$. It means that $S$ is negative and lies  in the region $-1\geq S < 0$.
 A negative $S$  is the reason for the term "reflective scattering". Its interpretation will be discussed in the next section.

The value of the collision energy corresponding to the complete absorption of the initial state
under the central collisions  $S(s,b)|_{b=0}=0$
is denoted as $s_r$ and the estimates for the value of $s_r$ are of order of  few  TeV \cite{srvalue}.
At the energies $s\leq s_r$ the scattering in the whole range of impact parameter variation
has a shadow nature (Fig. 1), it means that solution of the unitarity equation for the  elastic  amplitude has the form:
\begin{equation}\label{shad}
f(s,b)=\frac{1}{2}[1-\sqrt{1-4h_{inel}(s,b)}], 
\end{equation} 
which assumes a direct coupling of elastic scattering to multiparticle production often called shadow scattering.
When the  energy value becomes larger than 
 $s_r$, the scattering picture at small values of impact parameter 
($b\leq r(s)$, where  $S(s,b=r(s))=0$) starts to acquire a reflective contribution. At such energy    and impact parameter values  unitarity gives for the  elastic  amplitude  another form:
\begin{equation}\label{ashad}
f(s,b)=\frac{1}{2}[1+\sqrt{1-4h_{inel}(s,b)}].
\end{equation}
\begin{figure}[hbt]
	\vspace{-0.42cm}
	\hspace{-2cm}
%	\begin{center}
		\resizebox{16cm}{!}{\includegraphics{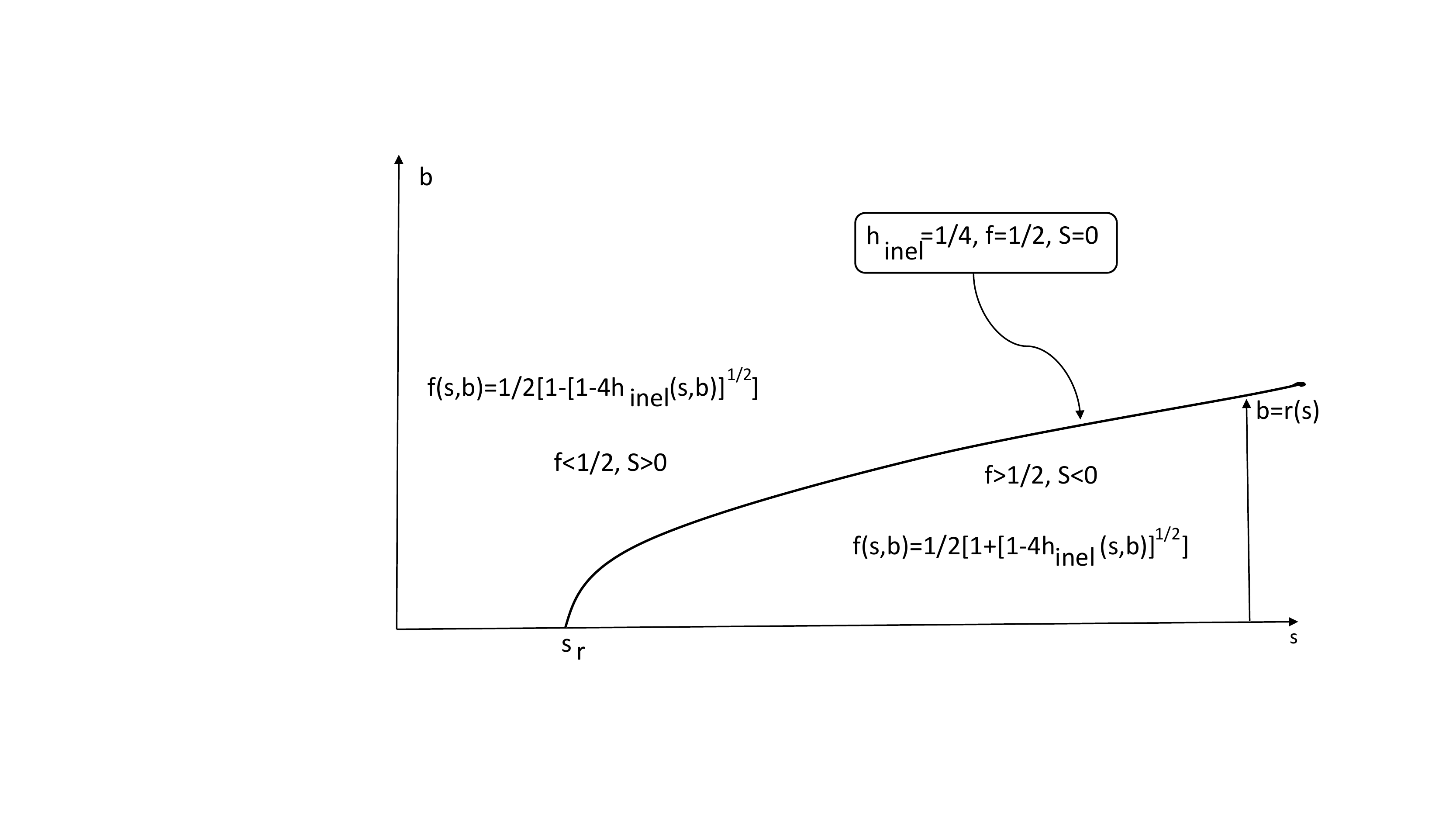}}		
		%\includegraphics{dsdb2n.eps}
%	\end{center}
	\vspace{-2cm}
	\caption{Schematic representation of the regions in $s$ and $b$ plane corresponding to absorptive ($S>0$) and reflective ($S<0$) scattering modes.}	
\end{figure}	 
Eq. (\ref{ashad}) corresponds to growing decoupling of the elastic scattering from multiparticle production. It can exist in the limited range of  the impact parameter values only, namely, at $b\leq r(s)$ since at larger values of $b$: $f\sim h_{inel}$. 
At $s>s_r$ the function $h_{inel}(s,b)$ has a peripheral $b$-dependence. Note, that
\begin{equation}
\label{der}
\frac{\partial h_{inel}(s,b)}{\partial b}=S(s,b)\frac{\partial f(s,b)}{\partial b},
\end{equation}
where $S(s,b)$ is negative at $s>s_r$ and $b< r(s)$.

\section{Reflection and color conductivity}
The  elastic scattering matrix element can be written in the form
\begin{equation}
S(s,b)=\kappa (s,b)\exp[2i\delta(s,b)].
\end{equation}
Here $\kappa $ and $\delta $ are the real functions and $\kappa$ can vary in the interval
$0\leq \kappa \leq 1$.  This function is called an absorption factor and its value $\kappa =0$ corresponds to complete absorption of the initial state, 
\begin{equation}
\kappa^2 (s,b)=1-4h_{inel}(s,b).
\end{equation}
The function $S(s,b)$ can be nonnegative in the whole region of the impact parameter variation or have negative values in the region $b<r(s)$ when the energy is high enough, i.e. at $s>s_r$. 
Under the reflective scattering, $f>1/2$, an increase of elastic scattering amplitude $f$ corresponds to decrease of $h_{inel}$ according to 
\[
(f-{1}/{2})^2={1}/{4}-h_{inel}
\]
and, therefore, the term antishadowing has  been initially used for description of such scattering mode emphasizing that the reflective scattering is correlated with the self-damping of the inelastic channels contribution \cite{baker} and increasing decoupling of the elastic scattering from multiparticle production dynamics.

Transition to the negative values of $S(s,b)$ means that the  phase $\delta$ changes its value from $0$ to $\pi/2$.  The term reflective is taken from optics
since the phases of incoming and outgoing  waves differ by $\pi$. It happens when the reflecting medium is optically denser (i.e. it has  a higher refractive index than the  medium where incoming wave travels before encounter  the scatterer). 
Thus, there is an analogy with the sign change  under reflection of the electromagnetic wave by  surface of a  conductor. This occurs because of  the  electromagnetic field  generates a current in the medium. 
The energy evolution of the scatterer   leads to appearance of  the reflective scattering mode if one admits the unitarity saturation in the limit of $s\to\infty$. 

Reflective scattering mode does not assume
any kind of hadron transparency during  the head-on collisions. Contrary, it is about the geometrical elasticity  (cf. \cite{usprd}).  The term transparency is relevant for the energy and impact parameter region responsible for the shadow scattering regime only, i.e. where $f<1/2$ (cf. Fig 1). The interpretation of the reflective scattering mode based   on the consideration of inelastic overlap function alone is, therefore, a deficient one. It could lead to an  incorrect rendering based on idea of the formation of the hollow fireball (e.g. filled by the disoriented chiral condensate) in the intermediate state of hadron--hadron interaction (cf. \cite{blr}) and consideration of  the central region as  the transparent one.

The emerging physical picture of   high energy  hadron interaction region  in transverse plane  can be visualized  then in a form of
 a reflecting
disk (with its albedo approaching to complete reflection at the center) which is surrounded by a   black ring 
(with complete absorption, $h_{inel}=1/4$) since the inelastic overlap function $h_{inel}$ has a prominent peripheral form  at $s\to\infty$ in this scattering mode.
The reflection  mode implies that the following  limiting behavior\footnote{Despite the limiting behavior of $S(s,b)$ corresponds to $S\to -1$ at $s\to\infty$ and fixed $b$, the gap survival probability, contrary to conclusion of  \cite{khoze}, tends to zero at $s\to\infty$ (cf. \cite{gsp}).}  $S(s,b)|_{b=0}\to -1$ will take place at $s\to\infty$. Of course, it is supposed a monotonic increase  of the amplitude $f$ with energy to its unitarity limit $f=1$, and an artificial option  of its nonmonotonic energy dependence at fixed values of $b$  is excluded\footnote{Such nonmonotonic behavior might result (after integration over $b$)  in peculiar distortions superimposed onto the rising  energy dependence of the total cross-section.}.

QCD is a  theory of hadron interactions  with colored objects confined inside those entities.  Thus, one can imagine that the color conducting medium is being formed instead of color insulating one when the energy of the interacting hadrons increases beyond some threshold value. Properties of such medium are under active studies in nuclear collisions, but color conducting phase can be generated in hadron interactions too. Therefore, one can try  to associate  appearance of the reflective scattering mode   with formation of the color conducting medium in the intermediate state of hadron interaction (cf. Fig. 2). Such idea was briefly mentioned   in  ref. \cite{blr}.
\begin{figure}[hbt]
	\vspace{-0.42cm}
	\hspace{-2cm}
	%	\begin{center}
	\resizebox{16cm}{!}{\includegraphics{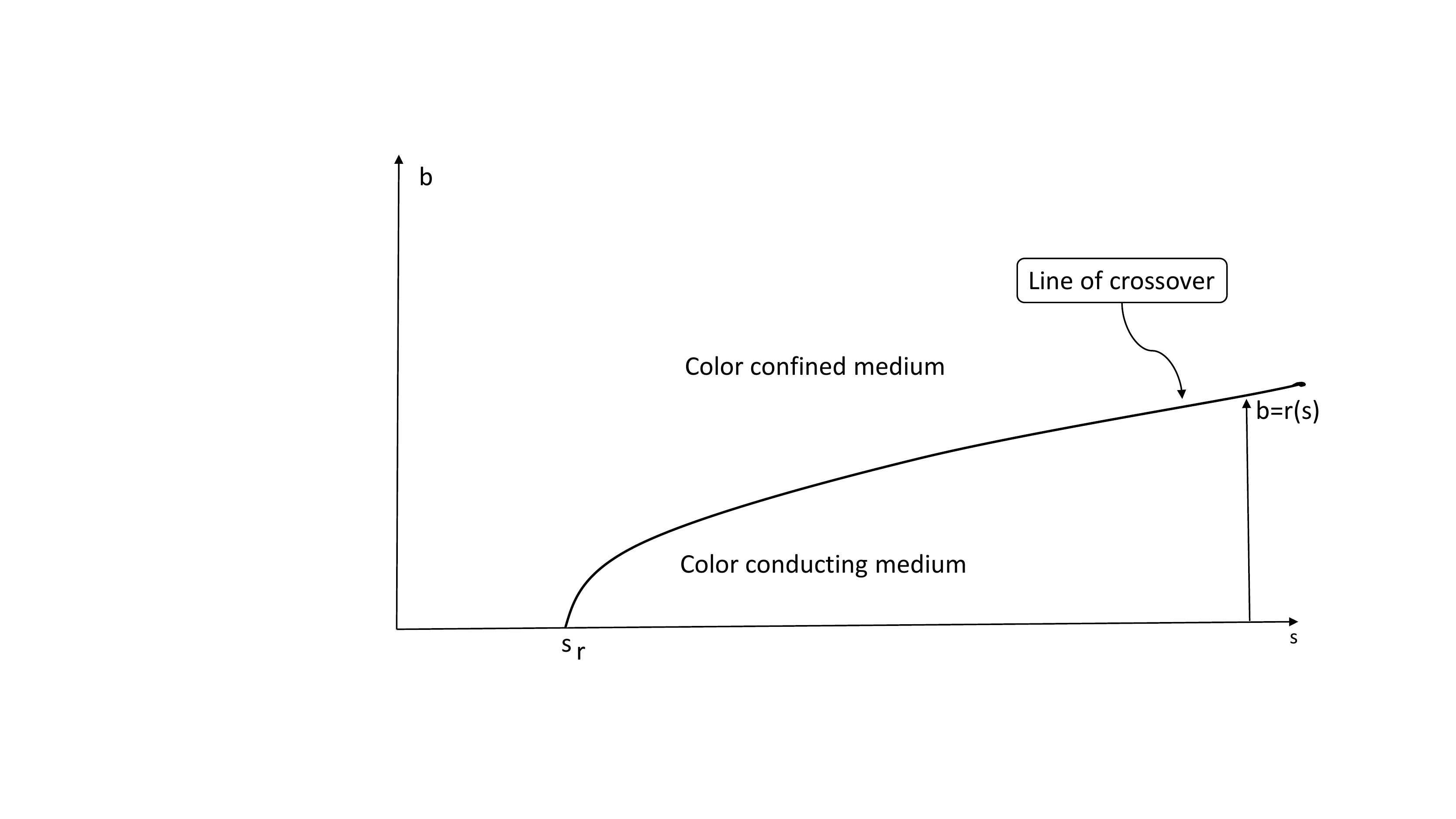}}		
	%\includegraphics{dsdb2n.eps}
	%	\end{center}
	\vspace{-2cm}
	\caption{Two phases of hadronic matter associated with the two respective scattering modes.}	
\end{figure}	 
The analogy is based  on replacement of an electromagnetic field  by a chromomagnetic field of QCD.  As  another example of a source for  analogy the phenomenon of Andreev reflection  at the boundary of the normal and superconducting phase applied to quark scattering at the interface between the cold quark-gluon phase and the color-superconductor can be pointed out \cite{and, sadz}.

 Formation of the color conducting medium in hadron collisions might also be responsible  for a number of collective effects such as correlations, anisotropic flows and others observed in small systems. Such effects can  arise  due to a ring-like shape  of the  impact--parameter  region responsible for the  multiparticle production processes. Such ring-like shape  is a result of the reflective scattering mode appearance. It implies  an important role of a coherent behavior of the deconfined matter formed under hadron interactions, resulting finally in the explicit collective effects  \cite{collect}.  

 We do not specify  the nature of color conducting medium, namely, there is no need to discuss what kind of constituents form it --- light point-like current colored  quarks or massive quasi-particles --- colored constituent quarks.  But, in any case one  can  expect  appearance of color conductivity in this medium under high-energy central hadron collisions.
 
 Indeed, the LHC  experiments  lead to an unexpected discovery of collective effects  in  small systems \cite{cms} (for  comprehensive list of references to the experimental results of ALICE, ATLAS, CMS and LHCb Collaborations cf. \cite{mangano} and for a brief review --- \cite{trib}).  A  prominent example is an observation of a "ridge" effect in two--particle  correlations in 
 $pp$-collisions in the events with high multiplicities \cite{weili}. 
 
 Interest to the collective effects is supported by  their relation to  dynamics of  quark-gluon plasma formation \cite{rafel}. And confinement of color is associated with collective, coherent interactions of quarks and gluons. It results in the formation of the asymptotic colorless states--hadrons. 
 
 The reflective scattering mode formation is consistent with the results of the impact parameter analysis at the LHC energy $\sqrt{s}=13$ \cite{alkin1}. This mode can be interpreted as a manifestation of the color conducting phase formation. The observation of the reflective scattering mode makes important the events classification depending on the  impact parameter of the collision. This mode  most significantly affects collisions with small impact parameters and hence it is important  to classify an impact parameter values of the particular hadron collision events. 
 
 It should  be emphasized that the collision geometry describes   the  hadron interaction region entirely  but not  the spacial properties of the  each participating hadron.
 
In what follows we consider definition of centrality  in small systems and the role of the reflective mode  at the LHC energy range.
 \section{Centrality in nucleus--nucleus collisions }
  The centrality  is  commonly accepted variable for description and classification of the collision events in nuclei interactions. This variable is related to the collision geometry-- degree of the collision peripherality. Thus, the centrality is given by the impact parameter value associated with  the general geometrical characteristics of a particular  collision event.
 
 As it was noted in \cite{olli} and \cite{olli1}, in experiments with nuclei (including hadron-nucleus reactions) the knowledge of centrality $c^N_b$ is extracted from either number of charged particles registered in the respective detector or the transverse energy measured in the calorimeter. Those quantities  are both denoted by $n$ and are relevant to  experimentally measurable quantity   $c^N$ also called centrality.   Superscript $N$ means nuclear collisions while superscript $h$ denotes pure hadron collisions. The definitions of $c^N_b$ and $c^N$  \cite{olli} are as follows
 \begin{equation}
 \label{centb}
 c_b^N\equiv \frac{\sigma^b_{inel}}{\sigma_{inel}},
 \end{equation}
 where 
 \[
 \sigma^b_{inel}=\int_0^bP^N_{inel}(b')2\pi b'db
 \]
 and  $P^N_{inel}(b)$ is probability distribution of the inelastic collisions  over the impact parameter $b$, while the experimentally measurable quantity 
 \begin{equation}
 \label{cent}
 c^N\equiv \int_n^\infty P^N(n')dn'.
 \end{equation}
 includes distribution over the multiplicity or the total transverse energy in the final state. 

 It should be noted that the energy dependence of the above quantities is tacitly implied and not indicated explicitly. The energy dependence, however, can be a nontrivial one in the collisions of nuclei as well as of hadrons, since size of interaction region, probabilities of interactions, multiplicities and transverse energies are the energy--dependent quantities in both cases. Evidently, the effects related to the energy dependence of all these quantities should be taken into account under analysis of the experimental data at the same value of centrality but at different energies.
 
 Under assumption that the probability $P^N(n)$ has a Gaussian   at a fixed value of the impact parameter $b$ \cite{bron},  the relation between  $c^N$ and $c_b^N$ has been obtained and discussed in \cite{olli}. The improvement of this reconstruction procedure with gamma distribution was performed in \cite{olli1}. This prescription allows one to extract a  knowledge on the impact parameter value from the experimental data independently of collision dynamics. Assumptions on the Gaussian  or gamma distributions are  general ones and do  not depend on the structure of the object under consideration, i.e. it could be applied equally  for nuclei and hadrons as well. Gamma distribution is preferable 
 in the regime where application of the central limit theorem is not justified \cite{olli1}. It seems that this approach to the impact parameter reconstruction can be used for hadron reactions also.

 It is important to emphasize again  that the proposed reconstruction of the impact parameter is not based on a particular nuclear interaction model and/or concept of participating nucleons. 
 
 \section{Reflective mode and centrality in hadron reactions}
 In view of the prominent collective effects observed in small systems, such as $pp$-collisions together with indications on the reflective scattering mode observation at the LHC, an introduction of centrality valid for any values of energy is useful    to classify the collision events. 
 
 The hadron scattering has  similarities as well as differences with the scattering of nuclei.  Geometrically, hadrons are the extended objects too, but a significant contribution to $pp$--interactions is provided by the elastic scattering with the ratio of elastic to total cross-sections $\sigma_{el}(s)/\sigma_{tot}(s)$  rising with energy. The elastic scattering of nuclei is not significant at high energies, nucleons are not confined in nucleus.  Hence, the geometrical characteristics of hadron collisions associated with the elastic scattering   are essential   for  the hadron dynamics, i.e. for the  development of QCD in its nonperturbative sector where the confinement  plays a crucial role. 
 
 It will be argued further that  definition of centrality based on the use of a straightforward analogy with nuclei interactions is not appropriate for the hadron interactions at the energies where the reflective scattering gives a significant contribution. To obtain a relevant universal definition, we  propose to use a full probability distribution $P^h_{tot}(s,b)$ in order to take into account the elastic channel events. The neglect of the elastic scattering events would lead to wrong estimation of centrality for the particular hadron collision.
 Thus, for the centrality $c_b^h(s,b)$ the following definition   is suggested
 \begin{equation}
 \label{centhb}
 c_b^h(s,b)\equiv \frac{\sigma^b_{tot}(s)}{\sigma_{tot}(s)},
 \end{equation}
 where
 \[
 \sigma^b_{tot}(s)=8\pi\int_0^b\mbox{Im}f(s,b') b'db'
 \]
 is the impact--parameter dependent cumulative contribution into the total cross--section, $\sigma^b_{tot}(s)\to \sigma_{tot}(s)$ at $b\to\infty$.  In Eq. (\ref{centhb}) the total (elastic plus inelastic) contribution replacing the inelastic cross--section only were used.
 
  It should be noted, that there is nothing wrong with definition of centrality in the form of
 Eq. (\ref{centb}) in case of hadron scattering at the energies where the reflecting scattering mode is not presented,
 but its presence  at higher energies changes form of an inelastic overlap function $h_{inel}(s,b)$ from a central to  peripheral one with maximum at $b\neq 0$. Therefore,  Eq. (\ref{centb}) for centrality in this case ceases to be valid. The use of centrality in this form  would lead to a distorted dependence when its value would not reflect real collision geometry. 
 
 Contrary to the inelastic overlap function $h_{inel}(s,b)$ , the function $\mbox{Im}f(s,b)$  at the LHC energies  has a central impact parameter profile with a maximum  located at $b=0$ \cite{alkin1}. 
 The amplitude $f(s,b)$ is the Fourier--Bessel transform of the scattering amplitude $F(s,t)$:
 \begin{equation}\label{imp}
 F(s,t)=\frac{s}{\pi^2}\int_0^\infty bdbf(s,b)J_0(b\sqrt{-t}).
 \end{equation}
 The definition  Eq. (\ref{centhb}) can be inverted, namely,
 one can consider centrality as an observable measured  in hadron collisions. Eq. (\ref{centhb}) then can be used for  restoration of  the elastic scattering amplitude, more specifically the function $\mbox{Im} f(s,b)$ can be calculated, if the impact parameter dependence of  $c_b^h(s,b)$ is experimentally known.
 The inverted relation corresponding to Eq. (\ref{centhb}) written in the differential form gives:
 \begin{equation}
 \label{centhbder}
 \mbox{Im} f(s,b)=\frac{\sigma_{tot}(s)}{8\pi b}\frac{\partial c_b^h(s,b)}{\partial b}.
 \end{equation}

 The impact parameter representation provides a simple semiclassical picture of hadron scattering, e.g. head--on or central collisions correspond to small impact parameter values.
 From Eq. (\ref{centhbder}) one can easily get the inequality
 \begin{equation}
 0\leq\frac{\partial c_b^h(s,b)}{\partial b}\leq \frac{8\pi b}{\sigma_{tot}(s)}
 \end{equation}
 or in the integral form
 \begin{equation}
 0\leq { c_b^h(s,b)}\leq \frac{4\pi b^2}{\sigma_{tot}(s)}
 \end{equation}
 for $b\leq R(s)$, $R(s)\sim \frac{1}{\mu} \ln s$, where $\mu$ is determined by the value of a pion mass.

 To demonstrate a transition to the reflective scattering mode explicitly we use the unitarization scheme which represents the scattering amplitude $f(s,b)$ in the rational form of one-to-one transform and allows its variation in the whole interval allowed by unitarity \cite{umat}.
 Respective form for the function $S(s,b)$ is written in this case as a known Cayley transform mapping nonnegative real numbers to the interval $ [-1, 1]$\footnote{This one-to-one transform maps upper 
 	half-- plane into a unit circle in case when $U$ and $S$ both are complex functions and the value of the function $S=0$ is reached at finite values of energy and impact parameter.} :
 \begin{equation}
 S(s,b)=\frac{1-U(s,b)}{1+U(s,b)}. \label{umi}
 \end{equation}
  It should be repeated here  that we neglect by the real part of the scattering amplitude $f(s,b)$.
 The real, nonnegative function $U(s,b)$ can be considered as an input or bare amplitude which is subject to the unitarization procedure. 
 The models of a different kind can be used for construction of a particular functional form of $U(s,b)$.
 The most of the models provide monotonically increasing
 dependence of the function $U(s,b)$ on energy (e.g. power-like one) and its exponential decrease with  the impact parameter (due to analyticity in the Lehmann-Martin ellipse).
 The value of the  energy corresponding to the complete absorption of the initial state
 at the central collisions  $S(s,b)|_{b=0}=0$
 is denoted as $s_r$ and  determined by the  equation
 $U(s_r,b)|_{b=0}=1$\footnote{The old (pre--LHC) numerical estimates  of $s_r$ have given for its value  $\sqrt{s_r}=2-3$ $TeV$ \cite{srvalue}.}.
 In the energy region $s\leq s_r$ the scattering in the whole range of impact parameter variation
 has a shadow nature and   high multiplicities  are associated with central collisions in the geometrical models. 

\section{Centrality and geometrical models}
 A wide class of the geometrical models (relevant for the centrality discussion) allows one to assume that $U(s,b)$ has a factorized form (cf. \cite{factor} and references therein):
 \begin{equation}\label{usb}
 U(s,b)=g(s)\omega(b),
 \end{equation}
 where $g(s)\sim s^\lambda$ at the large values of $s$, and the power dependence guarantees asymptotic growth of the total cross--section $\sigma_{tot}\sim \ln^2 s$. Such factorized form corresponds to a common source for the increase with energy of the total cross--sections and the slope of the diffraction cone in  elastic scattering. The particular simple form of the function $\omega(b)\sim \exp(-\mu b)$ has been chosen to meet the analytical properties of the scattering amplitude. This form of  $\omega(b)$ also assumed by the physical picture grounded on its
 representation as a convolution of the two energy--independent hadron pionic-type matter distributions (cf. Fig. 3, it illustrates  notion of centrality  in hadron scattering) in transverse plane:
 \begin{equation}
 \omega (b)\sim D_1\otimes D_2\equiv \int d {\bf b}_1 D_1({\bf b}_1)D_2({\bf b}-{\bf b}_1).
 \end{equation} 
 \begin{figure}[hbt]
 	\vspace{-0.4cm}
 	\hspace{-1cm}
 	%	\begin{center}
 	\resizebox{16cm}{!}{\includegraphics{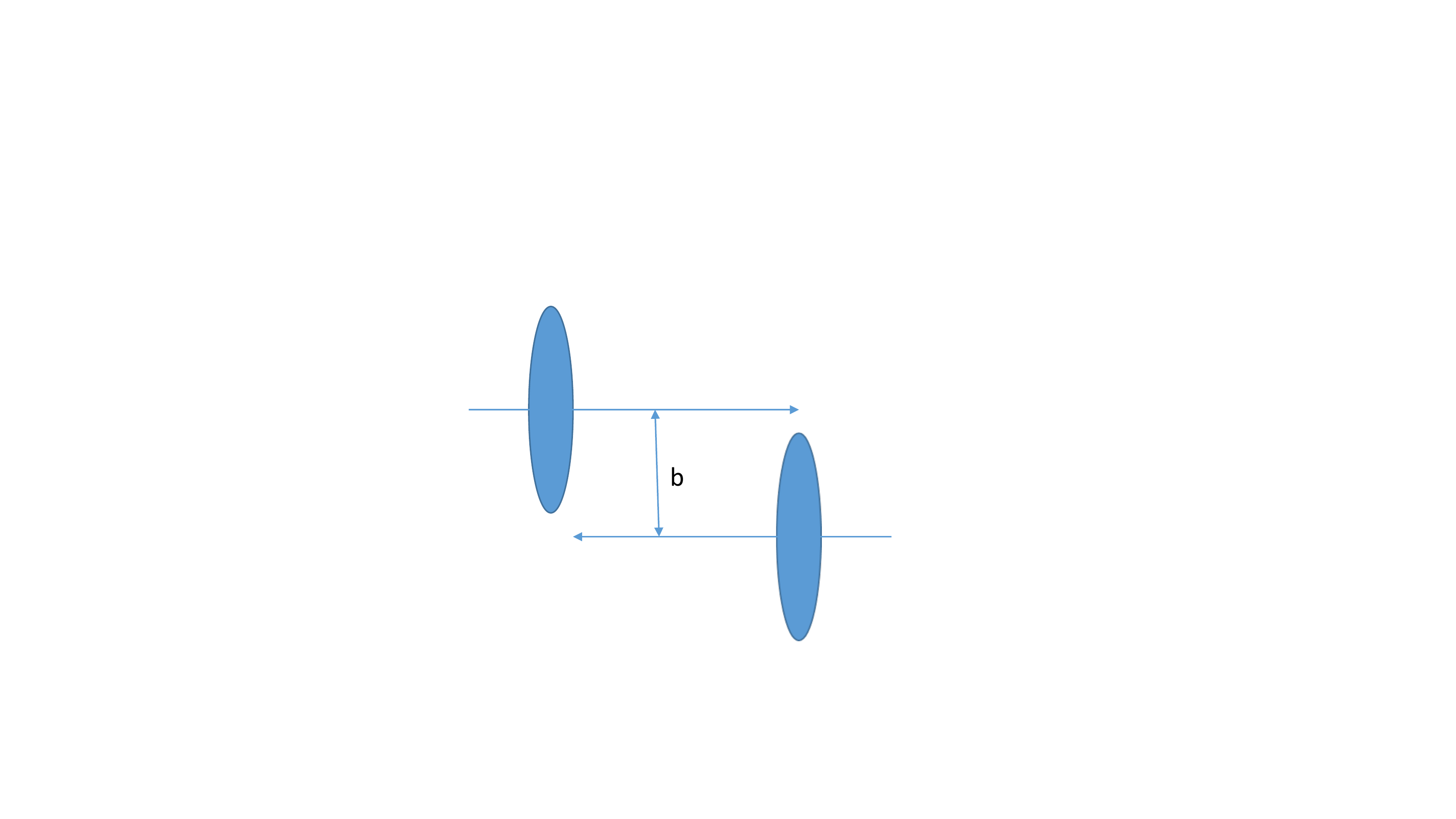}}		
 	%\includegraphics{dsdb2n.eps}
 	%	\end{center}
 	\vspace{-2cm}
 	\caption{Schematic view of  hadron scattering  with the  impact parameter $b$ in the geometric models (cf. e. g.  \cite{hei, chy, low,  sof}) .}	
 \end{figure}	 
 Parameter $\mu$ should be then equal  to the doubled value of a pion mass. 
 
 Of course, a  weak energy dependence of  centrality  can allow one to use it as a parameter for the data analysis at  different energies in hadron reactions.
 Indeed,  asymptotically, the centrality $c_b^h(s,b)$, defined according to Eq. (\ref{centhb}), decreases with energy slowly, like $1/\ln^2(s)$ at fixed impact parameter values. Unitarization generate its dependence 
 $\sim b^2/\ln^2(s)$ (since $f(s,b)$ saturates unitarity limit, i.e. $f(s,b)\to 1$ , at $s\to\infty$). Such slow energy decrease of centrality  allows one to compare the data at different energies and  approximately the same value of centrality provided     the energy values are high enough and are not too much different. Contrary, a strong energy dependence of centrality would bring problems under comparison of the data obtained at the same values of centrality  and different energies.

 To  justify further   use of the suggested form of Eq. (\ref{centhb}) for centrality instead of  Eq. (\ref{centb}), we examine Eq. (\ref{centb}) 
 aiming to construct a counterexample, i.e. to demonstrate that  such option is leading to the strong energy dependence of centrality defined that way, and therefore Eq. (\ref{centb})    is not an appropriate  definition  for small systems. 
 
 Thus, inelastic overlap function   $h_{inel}(s,b)$ in  Eq. (\ref{hinell})
 when the  function $U(s,b)$ is chosen in the form of Eq. (\ref{usb}) allows one to calculate  explicitly the centrality given by Eq. (\ref{centb}). In this case: 
 \begin{equation}
 \label{cbh}
 c_b^h(s,b) = \frac{1}{\ln(1+g(s))}\left[\ln\frac{1+g(s)}{1+g(s)\exp(-\mu b)}-\mu b f(s,b)\right],
 \end{equation}
 where 
 \begin{equation}
 \label{fsb}
 f(s,b)=\frac{g(s)\exp(-\mu b)}{1+g(s)\exp(-\mu b)}.
 \end{equation}
 It leads to conclusion that at $s\to \infty$ and fixed value of $b$ one would expect strong energy decrease of the function $c_b^h(s,b)$ under such option, i.e.
 \begin{equation}
 \label{asymp}
 c_b^h(s,b)\sim \frac{1}{s^\lambda\ln(s)}.
 \end{equation}
 Eq. {\ref{asymp} corresponds to the statement made in \cite{hprod}, where it has been shown that centrality defined in a straightforward analogy with the case of nuclei collisions  cannot serve as a measure of the impact parameter but this quantity is to be associated with the dynamics of the multiparticle production when the value of the impact parameter can variate in the narrow region around $b=r(s)$. In this region the absorption is maximal in case of the reflective scattering domination at $s\to\infty$ and centrality constructed in that way is just a cumulative contribution of the edge  (cf. \cite{edge}). Contrary, it is proposed to define centrality as an impact--parameter dependent cumulative contribution of all the interactions, elastic and inelastic ones, and it  has a central impact parameter profile corresponding to an intuitive expectation.

 An important problem is how to estimate  the impact parameter value in a given $pp$-collision event from the data. It is evident that the  event classification by multiplicity of the final state  is not relevant for that purpose since a contribution of the elastic channel is then almost neglected. Moreover, centrality defined in that way  has a strong energy dependence as it has been shown due to increasing peripherality of the probability distribution over impact parameter $P_{inel}^h(s,b)$ with rising energy. 
 
 The most relevant observable seems to be a sum of the transverse energies of the final state particles. We can assume, following \cite{olli} and \cite{olli1}, the  Gaussian or gamma distributions of the transverse energy for the fixed value of impact parameter $b$   and extend the conclusions of \cite{olli} and \cite{olli1} for nucleus-nucleus or proton-nucleus collisions to proton-proton collisions.  Namely, fitting the experimental data and using Bayes' theorem for conditional probability one can effectively reconstruct the distribution of the impact parameter for a given  value of centrality determined by the total transverse energy of all  final particles. Further  details of such reconstruction can be found in the papers \cite{olli} and \cite{olli1}. 
 
 It should also be noted that observation of the non--Gaussian elliptic flow fluctuations in PbPb collisions \cite{nong} makes use of gamma distribution preferable at the LHC energies since the degree of the initial--state spatial anisotropy resulting in the elliptic flow is correlated with the collision impact parameter value of the particular event.
 \section{Conclusion}
  
 The   features associated with a character of $b$--dependence of hadron interactions are considered in this note.
 The $b$-dependent differential quantities are much more sensitive to the interaction dynamics that the overintegrated over $b$ ones. They are responsible for a number of conclusions such as the  peripheral form of the inelastic overlap function and,  correspondingly, the central distribution of the elastic overlap function over the impact parameter. 
 
  As it was mentioned in the beginning of this note, the problem of the scattering amplitude  real part account has no final solution nowadays. 
  Its account  could, in principle, change the form of the inelastic overlap function since $h_{inel}$ transforms in this case into the sum $h_{inel}+(\mbox{Re} f)^2$ in the unitarity equation  \cite{pl}.  Despite such hypothetical possibility exists,  it is not consistent with the experiment. Namely, the results of the quantitative impact parameter analysis  performed in \cite{alkin1} {\it with account of the real part} are not in favor of such option.  The resulting form   of inelastic overlap function remains to be a peripheral one  and a relevant form   of the elastic overlap functions is not changed, it remains to be a central one. This  is in favor  of the real part  neglect at least under qualitative considerations and justifies the approximation by  an imaginary scattering amplitude. The vanishing role of the real part of the scattering amplitude account has been emphasized in \cite{drem}, too.  It also provides a hint on the unitarity saturation at the asymptotic energies.

There were proposed a rendering for the reflective scattering mode based on formation of color--conducting medium in the intermediate state and  definition of centrality has been given  with account of this mode existence in small systems like $pp$--collisions.  The use of the transverse energy measurements in a calorimeter seems to be a more relevant method for centrality estimations  than the method based on  multiplicity measurements. The use of transverse energy measurements is a more universal method since it includes the  case  of  unitarity saturation.  

Nowadays, ATLAS and CMS experiments at the LHC are indeed using transverse energy measurements for centrality determination, but in the collisions of nuclei only. Experimental feasibility of centrality measurements  in case of hadron reactions needs to  be considered also. 
 
 Centrality extracted from the experimental data can be used for the elastic scattering amplitude reconstruction in the impact parameter space according to Eq. (\ref{centhbder}). The magnitude of this amplitude at small values of $b$ is essential for reflective scattering mode detection. It is due to the fact that dynamics of elastic $pp$--scattering  is described by  complex function $F(s,t)$ of the two Mandelstam variables $s$ and $t$.
  Any quantities integrated over $b$  (i.e. those taken at $-t=0$) are not sensitive to the details of their dependencies on  $b$ and/or $t$, respectively, and therefore they cannot provide required information relevant for the available accelerator energies. Much higher energies are needed for the possibility of making definite conclusions on the new scattering mode appearance if one  proceeds from the overintegrated quantities  only.
 
 The relation of centrality and $b$-space elastic amplitude   is  similar to the optical theorem (it relates the elastic scattering amplitude with  properties of all the collisions including elastic and inelastic ones).
 
 Evidently, the important issue  is further search for the unambiguous experimental manifestations of the reflective  mode in elastic and inelastic hadron interactions.  The centrality measurements in $pp$-collisions seems to be a promising way for that purpose since it would help to extract information on the typical impact parameter values of a particular collision event.

\section*{Acknowledgements}
We are grateful to E. Martynov for  the interesting discussions.

\end{document}